\documentclass[12pt]{spieman}  % 12pt font required by SPIE;
\usepackage{amsmath,amsfonts,amssymb}
\usepackage{graphicx}
\usepackage{setspace}
\usepackage{tocloft}
\usepackage{lineno}
\newcommand*\sunrise{\textsc{Sunrise}}
\newcommand*\sunriseIII{\textsc{Sunrise III}}
\usepackage{color}
\newcommand\lr[1]{{\color{black}\rm#1}}

\title{High-speed data processing onboard sunrise chromospheric infrared spectropolarimeter for the {\sunriseIII} balloon telescope} 

\author[a.*]{Masahito Kubo}
\author[a]{Yukio Katsukawa}
\author[b]{David Hern{\'a}ndez Exp{\'o}sito}
\author[c]{Antonio S{\'a}nchez G{\'o}mez}
\author[c]{Mar{\'\i}a Balaguer Jimen{\'e}z}
\author[c]{David Orozco Su{\'a}rez}
\author[c]{Jos{\'e} M. Morales Fern{\'a}ndez} 
\author[c]{Beatriz Aparicio del Moral}
\author[c]{Antonio J. Moreno Mantas}
\author[c]{Eduardo Bail{\'o}n Mart\'{i}nez}
\author[c]{Jose Carlos del Toro Iniesta}
\author[a]{Yusuke Kawabata}
\author[b,d]{Carlos Quintero Noda}
\author[a]{Takayoshi Oba}
\author[a]{Ryohtaroh T. Ishikawa}
\author[e]{Toshifumi Shimizu}

\affil[a]{National Astronomical Observatory of Japan, 2-21-1 Osawa, Mitaka, Tokyo, Japan, 181-8588}
\affil[b]{Instituto de Astrof\'{i}sica de Canarias, C. V\'{i}a L{\'a}ctea, s/n, 38205, San Crist{\'o}bal de La Laguna, Santa Cruz de Tenerife, Spain}
\affil[c]{Instituto de Astrof\'{i}sica de Andaluc\'{i}a, Glorieta de la Astronom\'{i}a s/n, 18008, Granada, Spain}
\affil[d]{Departamento de Astrof\'isica, Univ. de La Laguna, La Laguna, Tenerife, E-38205, Spain}
\affil[e]{Institute of Space and Astronautical Science, Japan Aerospace Exploration Agency, 3-1-1 Yoshinodai, Chuo, Sagamihara, Kanagawa, Japan, 252-5210}

\cftpagenumbersoff{figure}
\cftpagenumbersoff{table} 
\begin{document} 
\maketitle

\begin{abstract}
The Sunrise Chromospheric Infrared spectroPolarimeter (SCIP) has been developed for the third flight of the {\sunrise} balloon-borne stratospheric solar observatory.
The aim of SCIP is to reveal the evolution of three-dimensional magnetic fields in the solar photosphere and chromosphere \lr{using} spectropolarimetric measurements with a polarimetric precision of 0.03\% (1$\sigma$).
Multiple lines in the 770 and 850 nm wavelength bands are simultaneously observed with two 2k$\times$2k CMOS cameras at a frame rate of 31.25 Hz.
Stokes profiles are calculated onboard by accumulating the images modulated by a polarization modulation unit, and then compression processes are applied to the two-dimensional maps of the Stokes profiles.
This onboard data processing effectively reduces the data rate.
SCIP electronics can handle large data formats at high speed.
Before the implementation into the flight SCIP electronics, a performance verification of the onboard data processing was performed with synthetic SCIP data that were produced with a numerical simulation modeling the solar atmospheres.
Finally, we verified that the high-speed onboard data processing was realized on ground with the flight hardware by using images illuminated by natural sunlight or an LED.
\end{abstract}

% Include a list of up to six keywords after the abstract

% Include email contact information for corresponding author
{\noindent \footnotesize\textbf{*}Masahito Kubo,  \linkable{masahito.kubo@nao.ac.jp} }

\begin{spacing}{2}   % use double spacing for rest of manuscript

\section{Introduction}
\label{sec:intro}  % \label{} allows reference to this section
The {\sunrise} balloon-borne stratospheric solar observatory is an international project to observe the Sun at a high spatial resolution with a 1-m telescope during balloon flight\cite{2011SoPh..268....1B, 2017ApJS..229....2S}.
The flight altitude is \lr{$\sim$35 km} above the Atlantic Ocean and the flight time is more than \lr{5} days from Sweden to Canada.
The Sunrise Chromospheric Infrared spectroPolarimeter (SCIP)\cite{2020SPIE11447E..0YK} is a slit-scanning spectropolarimeter, and one of the three focal plane instruments\cite{2020SPIE11447E..AKF, 2022SPIE12184E..2GA} developed for the third flight of {\sunrise}.
\lr{The} SCIP runs multi-wavelength spectropolarimetric observations with a polarimetric precision of 0.03\% (1$\sigma$) and a spatial resolution of 0.21 arcsec\cite{2020SPIE11447E..AJT, 2020SPIE11447E..ABU, 2022SPIE12184E..27K,2022SPIE12184E..2BK}.
The spatial resolution of \lr{the} SCIP corresponds to the diffraction limit of \lr{1 m} for the telescope at the wavelength of 850 nm.
Two orthogonal linearly polarized beams are spatially separated in the direction perpendicular to the spectral dispersion by a polarization beam-splitter and simultaneously recorded by two spectropolarimeter (SP) cameras\cite{2023FrASS..1067540O}: SP1 camera for Ca II  lines in the 850 nm wavelength band and SP2 camera for K I lines in the 770 nm wavelength band.
In addition to these two SP channels, a slit-jaw (SJ) imager observes two-dimensional solar images on the slit.
The onboard data processing, such as demodulation, bit compression, and image compression, significantly reduces the data rate for spectropolarimetric observations.
The usefulness of the onboard data processing for space-borne spectropolarimetry has been successfully demonstrated by the spectropolarimeter\cite{2013SoPh..283..579L} of the Solar Optical Telescope (SOT)\cite{2008SoPh..249..167T} on the Hinode mission\cite{2007SoPh..243....3K}.
An advancement over the Hinode/SOT spectropolarimeter is that the onboard data processing of \lr{the} SCIP can handle images from cameras at a data rate 100 times higher than that of the Hinode/SOT spectropolarimeter.
The size of the read-out frame is 1024$\times$112 pixels for each of the two CCD sensors for the Hinode/SOT spectropolarimeter and 2048$\times$2048 pixels for each of the two CMOS sensors for \lr{the} SCIP.
The frame rate is 10 Hz for the Hinode/SOT spectropolarimeter and 31.25 Hz for \lr{the} SCIP.
The polarization modulation is done by continuously rotating waveplates in both instruments, implying that the camera is read continuously during the polarization modulation.
A larger data format is often requested for cameras because a larger field-of-view and wider wavelength coverage can provide deeper scientific insights.
Particularly for SCIP, the orthogonal polarization of many Zeeman-sensitive absorption lines is measured by one camera to obtain the three-dimensional magnetic field structures from the photosphere to chromosphere\cite{2017MNRAS.464.4534Q, 2017MNRAS.472..727Q, 2019MNRAS.486.4203Q}. 
Moreover, a fast polarization modulation is essential to \lr{detecting} the temporal evolution of magnetic field structures related to the dynamical chromospheric phenomena.
%For example, at least 15s cadence is needed if we try to trace the plasma moving within one resolution element at the speed of the sound in the chromosphere (10 km/s).
This fast polarization modulation requires a fast modulator and \lr{a} camera readout with a high frame rate.
Thus, the high-speed data processing is important for state-of-art spectropolarimeters.

\subsection{Observation Mode}
\label{sec:obs_mode}

\begin{figure}
\begin{center}
\begin{tabular}{c}
\includegraphics[height=12cm]{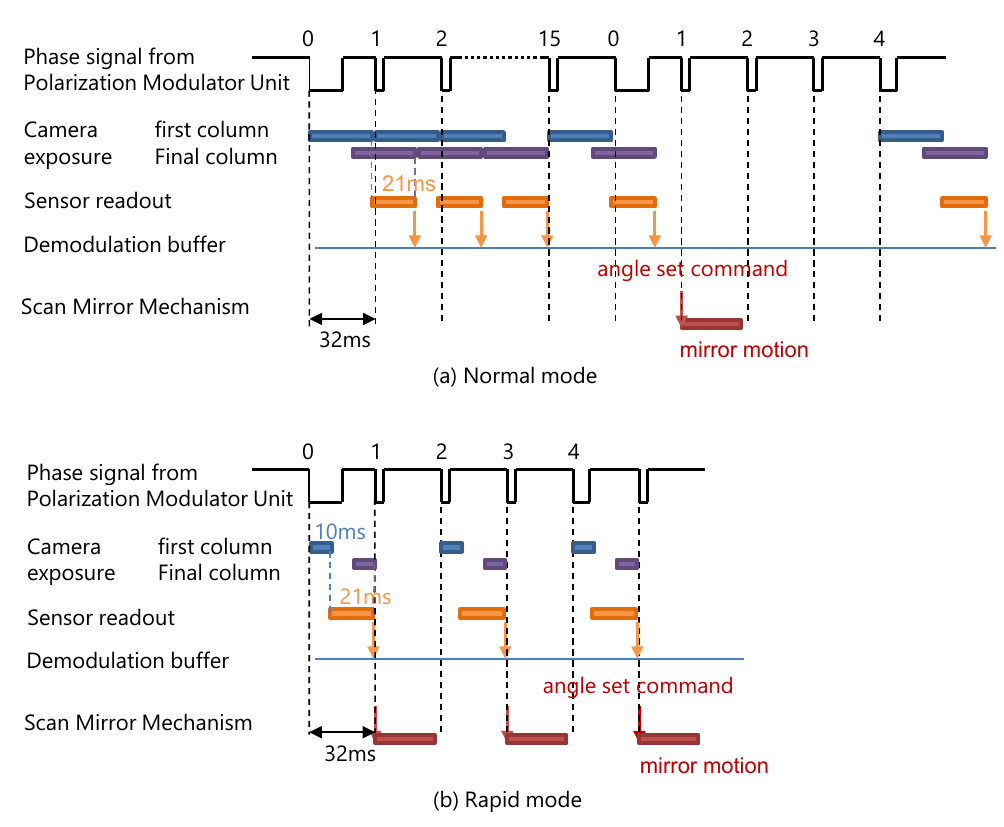}
\end{tabular}
\end{center}
\caption 
{ \label{fig:obsmode}
Control sequence of observations in \lr{(a) normal mode and (b) rapid mode.}} 
\end{figure} 

Two observing  modes are available for \lr{the} SCIP, as shown in Fig.~\ref{fig:obsmode}.
The onboard data processing, camera exposures, and \lr{scan mirror mechanism} (SMM)\cite{2022SoPh..297..114O, 2020SPIE11445E..4FO} synchronously co-operate with the \lr{polarization modulation unit} (PMU)\cite{2020SPIE11447E..A3K, 2015SoPh..290.3081I} in the observing sequence.
The PMU continuously rotates a pair of waveplates at a constant rate of 0.512 s/rotation and sends a pulse (``\lr{phase} signal from Polarization Modulator Unit'' in Fig.~\ref{fig:obsmode}) every 22.5 \lr{deg} to the SCIP electronics unit.
The length of the pulse represents the phase of the PMU rotation: the longest, mid-long, and nominal pulses arrive every \lr{360 deg, 90 deg, and 22.5 deg}, respectively.
The camera \lr{starts} an exposure of the first column in its sensor at the leading edge of the pulses and then sends images to the \lr{data processing unit} (DPU), which consists of a system controller and a frame grabber, in the SCIP electronics unit.

A set of Stokes I, Q, U, V, and R states are produced in \lr{normal} mode.
\lr{The Stokes parameters represent the intensity (Stokes I), linear polarization (Stokes Q and U), and circular polarization (Stokes V) of light.}
The R state is used for the polarization calibration of Stokes V (see \lr{Sec}~\ref{sec:demodulation}).
The size of the output is 2048$\times$2048 pixels in the SP1 (850 nm) and SP2 (770 nm) cameras \lr{and} 640$\times$640 pixels in the SJ camera.
The field-of-view of SJ is 60 arcsec $\times$ 60 arcsec with a pixel sampling of 0.09 arcsec, and the slit length, which is \lr{the same} on the two SP channels, is 58 arcsec.
The frame rate is 16 exposures per one PMU rotation\lr{,} i.e., 32 ms per frame at a constant rate.
The acquisition of a set of images starts at the first longest or mid-long pulse, followed by their demodulation by DPU.
The shortest accumulation is 32 images during two PMU rotations (i.e., 1.024 s), and the largest accumulation is 640 images during 40 PMU rotations (i.e., 20.48 s).
A requirement on the polarization sensitivity for \lr{the} SCIP is 0.1\% (1$\sigma$) at 1.024 s integration and 0.03\% (1$\sigma$) at 10.24 s integration.
The scan mirror is moved to the next position at the edge of the PMU pulse just after the camera sensor readout.

In \lr{rapid mode}, only Stokes I is obtained at eight frames per one PMU rotation, i.e., every 64 ms, for the SP1 and SP2 cameras. 
The full field-of-view (58 arcsec $\times$ 58 arcsec) can be covered in approximately 40 s by the slit-scanning observations in \lr{rapid} mode.
The size of the sensor readout is 2048$\times$2048 and 2048$\times$100 pixels for the SP1 and SP2 cameras, respectively.
The size in the wavelength direction of the SP2 camera is kept small to reduce the data rate, which is limited by a \lr{gigabit ethernet} connection with a data storage in the \lr{instrument control system} (ICS).
The size of 100 pixels in the wavelength direction covers the most important line (K1 D1) in the 770 nm wavelength band. 
The number of pixels for the SP2 camera can be extended if \lr{a} higher data rate is allowed.
For the SJ camera, the image size is the same as that in the \lr{normal} model\r{,} but the frame rate is eight images per one PMU rotation.
The images of SP/SJ cameras are taken only in the even pulse of PMU, and the motion of SMM is conducted in the odd pulse, \lr{with the longest pulse of the phase signal from PMU being defined as the 0th pulse}.

\subsection{Onboard Processing}
\label{sec:onboard_porc}
\begin{figure}
\begin{center}
\begin{tabular}{c}
\includegraphics[height=16cm]{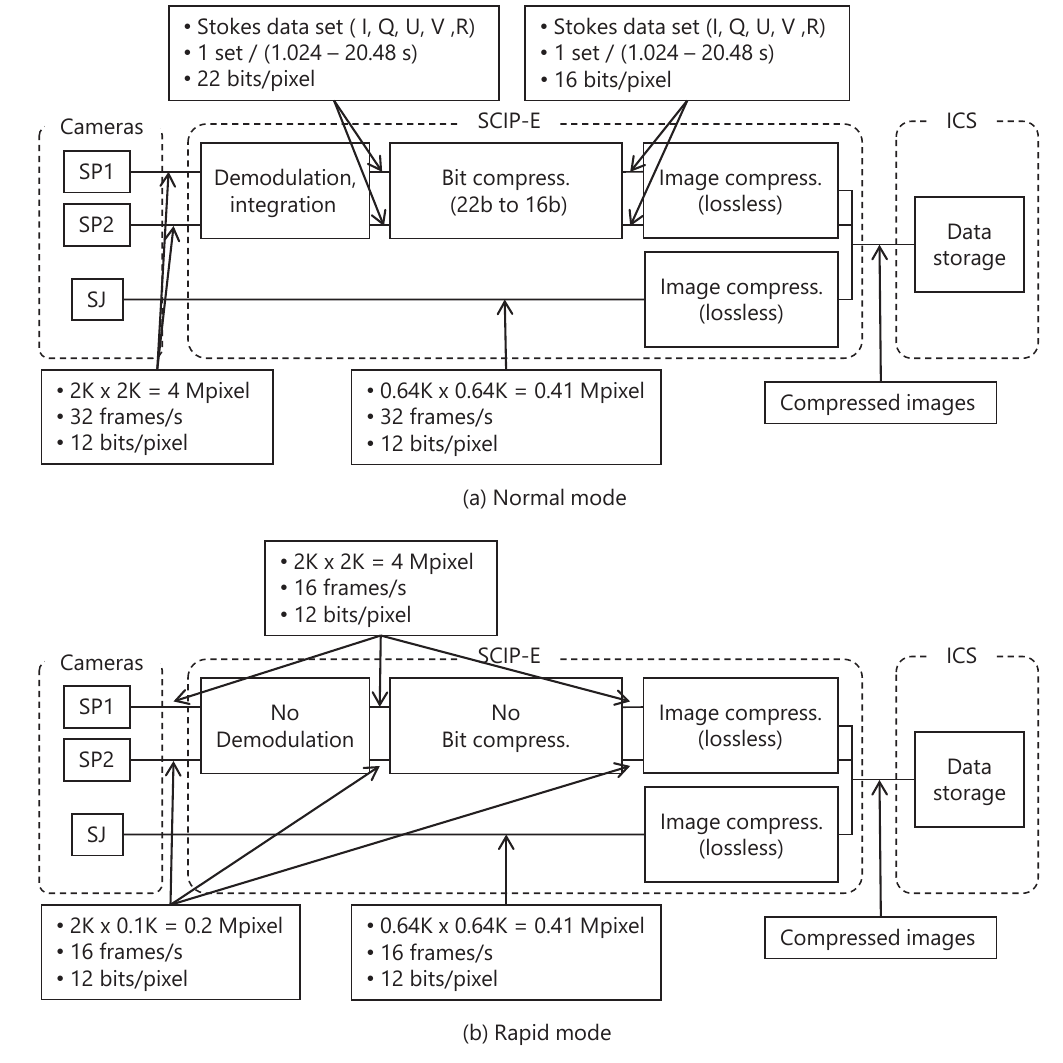}
\end{tabular}
\end{center}
\caption 
{ \label{fig:dataflow}
Onboard data flows in \lr{(a) normal mode and (b) rapid mode.}} 
\end{figure}

Figure~\ref{fig:dataflow} shows onboard data flows for \lr{the} SCIP.
Demodulation (\lr{Sec.}~\ref{sec:demodulation}), bit compression (\lr{Sec.}~\ref{sec:bit_compression}), and image compression (\lr{Sec.}~\ref{sec:image_compression}) processes are carried out for the images of the SP1 and SP2 cameras in the \lr{normal} mode.
Only image compression is processed for the SP cameras in the \lr{rapid} mode.
Demodulation and bit compression are always skipped for the outputs from the SJ camera, but image compression is applied.
These onboard processes are applied in the same way to both science and calibration data.
The SCIP data are sent to the data storage in ICS.
The data rate from the SCIP electronics unit should satisfy a \lr{gigabit ethernet} connection with the data storage in ICS.
The recorded data in the data storage will be recovered after the balloon flight. 
After that, the calibration of the science data will be performed on the ground to make real Stokes data.

The onboard processing is carried out in the frame grabber within the SCIP DPU. This frame grabber is a custom design implemented on a \lr{field-programmable gate array} (FPGA) provided by Xilinx, the Kintex Ultrascale XCKU040. The processing pipeline is programmed in a combination of custom cores and third-party cores. The custom cores are described in the VHDL language to implement the main functionalities of the data pipeline, i.e., the demodulation \lr{(integration), bit-compression and image compression} cores. The third-party cores are supplied by Xilinx in the Vivado development tool and are mainly responsible for the data movement between processing cores, external memory access and fixed-point mathematical operations (adder, multiplier\lr{,} and square root). In addition, the frame grabber is equipped with an embedded soft processor Microblaze, which is also provided by Xilinx. It is programmed in C.

\section{Onboard Demodulation}
\label{sec:demodulation}
\begin{figure}
\begin{center}
\begin{tabular}{c}
\includegraphics[bb=0 0 560 260, width=16cm,clip=]{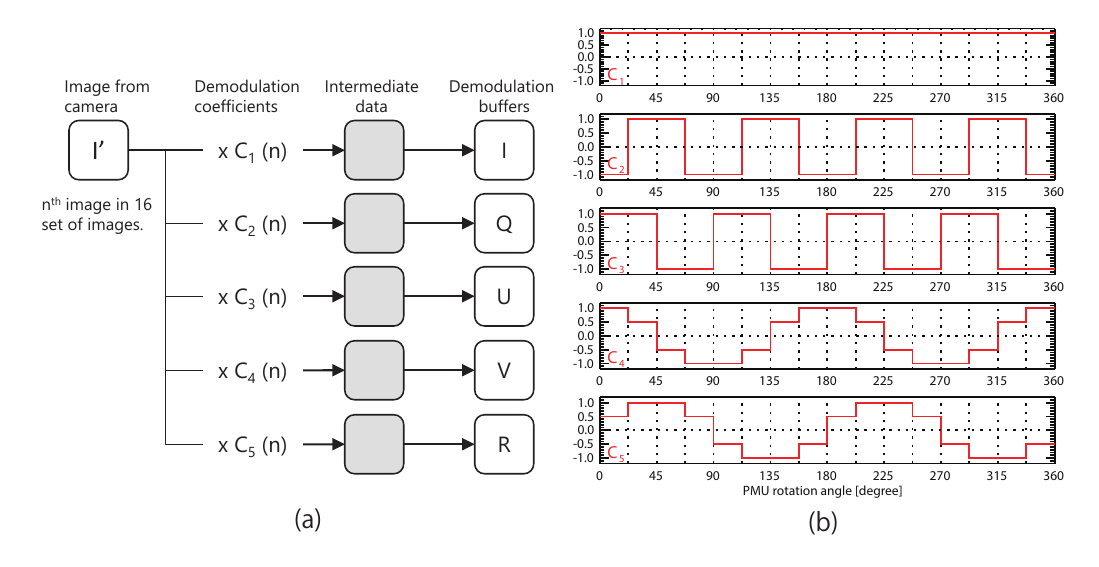}

\end{tabular}
\end{center}
\caption 
{ \label{fig:demodulation}
\lr{(a) Demodulation scheme and (b) demodulation coefficients. }} 
\end{figure}

A scheme of the integration and demodulation is shown in Fig.~\ref{fig:demodulation}.
In the \lr{normal} mode, the camera takes an image every 22.5 \lr{deg} of the PMU rotation. 
This image shows the modulated intensity, which is a combination of Stokes parameters multiplied by a factor corresponding to the phase of the PMU rotation. 
Therefore, the Stokes parameters can be obtained by accumulating multiple images taken at different PMU phases in the demodulation buffers. 
Before accumulation, the image taken at one phase of the PMU rotation is multiplied by a demodulation coefficient (``\lr{intermediate} data'' in Fig.~\ref{fig:demodulation}).
The demodulation coefficient varies among -1, -1/2, 1/2, and 1 according to the phase of the PMU rotation.
As a minimum dataset, \lr{the} SCIP is designed to calculate the Stokes parameters from 16 images obtained during one PMU rotation.

The CMOS sensor readout is not performed for all parts of the image simultaneously, but each column is read out in parallel along the wavelength direction. A new parameter R is introduced to compensate for this rolling shutter effect of the sensor\cite{2022ApOpt..61.9716K}. The R-parameter is defined to have a phase shift of 45 \lr{deg} with respect to Stokes V.  The speed of the rolling shutter is 0.010 ms/column\lr{,} and it produces a gap of about 21 ms between the first and \lr{final} column along the wavelength direction. This time gap corresponds to the phase shift of the PMU rotation up to 15 \lr{deg}. The phase shift of the PMU causes a crosstalk between Stokes Q and U and between Stokes V and R-parameter. The rolling shutter is  precisely controlled at a constant speed, \lr{so} the phase shift due to the rolling shutter can be accurately known. This allows us to apply the same demodulation matrix to all pixels of the image. The Stokes V is corrected \lr{using} the R-parameter with the phase shift due to the rolling shutter.

\begin{figure}
\begin{center}
\begin{tabular}{c}
\includegraphics[bb=0 0 425 481, height=17cm,clip=]{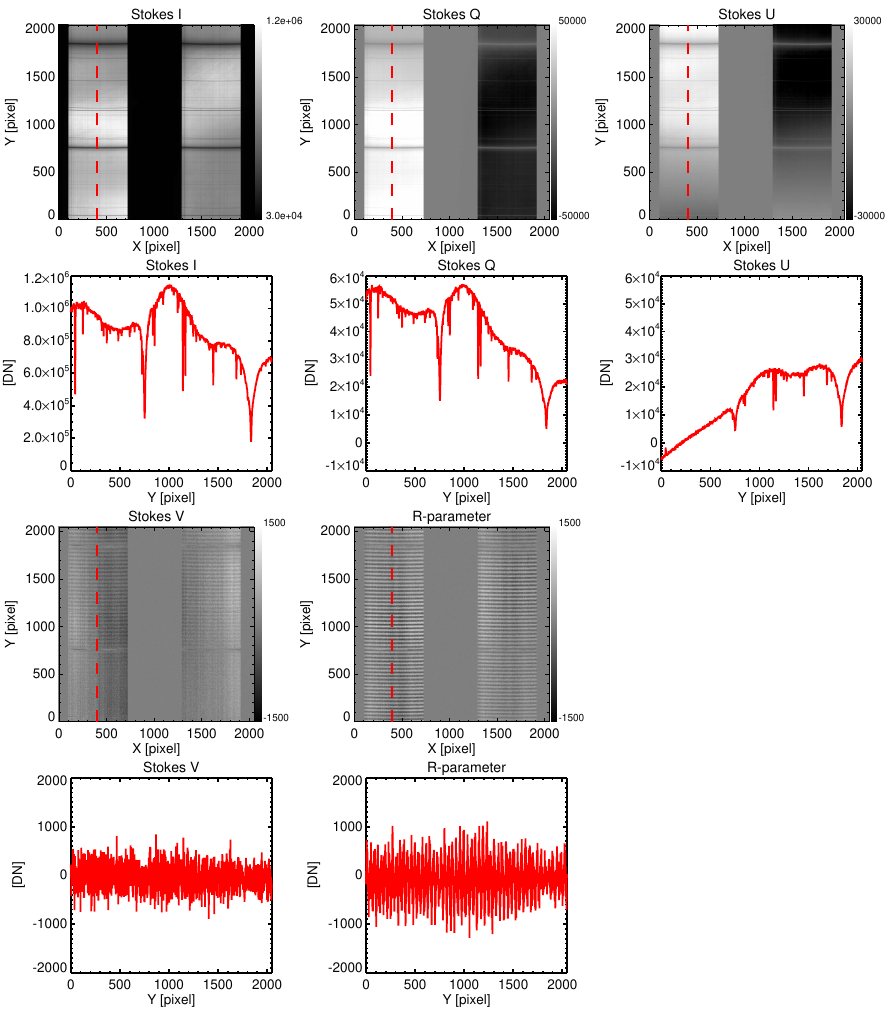}
\end{tabular}
\end{center}
\caption 
{ \label{fig:SP1_obs}
Example dataset of the SP1 channel (850 nm) observed in the \lr{normal} mode. 
The panels in the first and third rows show the results of the onboard demodulation. 
The horizontal and vertical axes represent the spatial and wavelength directions, respectively.
The panels in the second and fourth rows show the Stokes I, Q, U, V, and R profiles along the vertical dashed line in their upper panels.
} 
\end{figure}

SCIP can output raw data with the onboard demodulation disabled.
For the verification of the onboard demodulation, known polarized light was fed into the SCIP optical unit by a test optical system\cite{2022ApOpt..61.9716K}.
The \lr{datasets} with and without the onboard demodulation were sequentially obtained, and we confirmed that they were consistent with each other. 
An example of the data after the onboard demodulation is shown in Fig.~\ref{fig:SP1_obs}. 
This dataset was obtained on ground with the SP1 channel when natural sunlight was fed into the SCIP optical unit through the {\sunriseIII} telescope after mounting it on the telescope. 
%The horizontal and vertical axis represents the spatial and wavelength direction, respectively.
The two illuminated areas correspond to the orthogonal polarization beams.
The most prominent horizontal dark lines in the Stokes I map correspond to the Ca II 854.2 and 849.8 nm lines.
Relatively large linear polarizations (Stokes Q and U) are induced by folding mirrors in the telescope and light-distribution optics located upstream of the SCIP optical unit.
The decrease in Stokes Q and increase in Stokes U along the wavelength direction is because of the rolling shutter of the CMOS sensor. 
The instrumental polarization and the effect of the rolling shutter can be calibrated using the polarization calibration data precisely measured before the flight\cite{2022ApOpt..61.9716K}.

%\begin{figure}
%\begin{minipage}[c]{0.5\hsize}
%\centering
%\includegraphics[width=8.5cm]{JATIS2022_demodulation.eps}
%\end{minipage}
%\begin{minipage}[c]{0.5\hsize}
%\centering
%\includegraphics[width=8.5cm]{JATIS2022_demodulation_co.ps}
%\end{minipage}
%\caption 
%{ \label{fig:demodulation}
%Demodulation scheme (left) and demodulation coefficients (right). } 
%\end{figure} 

\section{Bit Compression}
\label{sec:bit_compression}

\subsection{Bit-Compression Algorithm}
\label{sec:bit_compression_algorithm}
The raw images are provided from the cameras in \lr{a} 12-bit word format.
In the \lr{normal} mode, the maximum accumulation time in the onboard demodulation procedure is 20.48 s.
Considering the demodulation coefficients, the number of bits needed after the demodulation is unsigned 22 bits for Stokes I and signed 21 bits for Stokes \lr{Q, U, V,} and R-parameter. 
However, the input range to the standard image compression algorithm described in \lr{Sec.}~\ref{sec:image_compression_algorithm} is 16 bits.
Thus, it is necessary to perform bit compression to 16 bits before the image compression. 

A nonlinear image transformation based on a square root function, similar to that used in the Hinode/SOT spectropolarimeter\cite{2013SoPh..283..579L}, is employed. 
The bit-compression algorithm is described by the following equations:
\begin{eqnarray}
X &=& N \hspace{5.4cm}    (N \leq N_c) \\
X &=& round(a + \sqrt{bN + c}) \hspace{2cm}  (N \textgreater N_c), 
\end{eqnarray}
\lr{where $N$ is the input pixel data to the bit compression process, $X$ is the value after bit compression, $N_{c}$ is the boundary between linear and square root compression regions, and $a$, $b$, and $c$ are the constant values.}

The bit-compression parameters (\lr{$a$, $b$, and $c$}) are calculated to satisfy the following three equations with a given $N_c$:
\begin{eqnarray}
N_c &=& a + \sqrt{bN_c + c}, \\
dX/dN &=& 1 \hspace{1cm} at~N = N_c, \\
M_{max} &=& a + \sqrt{bN_{max} + c},
\end{eqnarray}
where M$_{max}$ is 2$^{16}$ counts.
The linear range ($N_c$) should be chosen to obtain compression errors \lr{that are} as small as possible. 
The photon noise for 20 s integration (up to 22 bits) is \lr{$\sim$}470 DN (1$\sigma$), considering the camera conversion factor (gain) of 0.052 DN/$e^{-}$.
We choose the linear range to be approximately three times the photon noise.

\begin{figure}
\begin{center}
\begin{tabular}{c}
\includegraphics[bb=0 0 510 170, height=5cm]{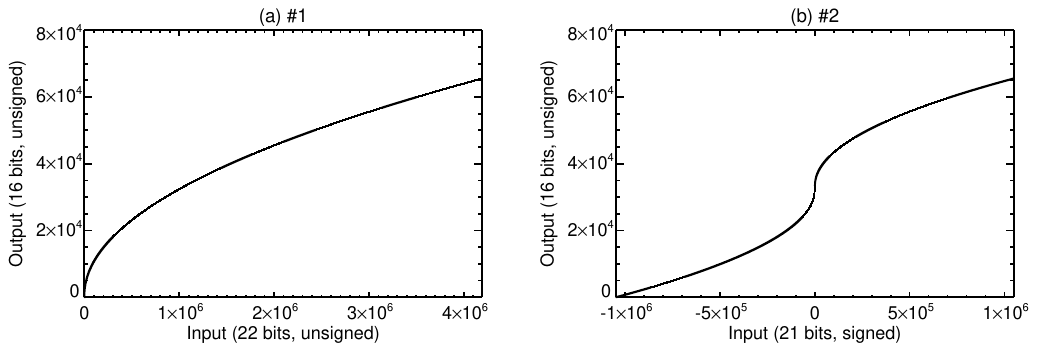}
\end{tabular}
\end{center}
\caption 
{ \label{fig:LUT}
Bit-compressed output after applying (a) for Stokes I and (b) for Stokes QUV and \lr{the} R-parameter.
} 
\end{figure}

The bit-compression process was previously implemented through look-up tables (LUTs) in the Hinode case because of the limitation on the processing capabilities.  
In this method, a LUT is computed once for the entire input range and stored in a memory. 
The input pixel of 22 bits for \lr{the} SCIP requires \lr{a} memory usage of \lr{$\sim$}8 MB for each LUT, which is 85 times larger than that in the Hinode case.
The internal memory resources on the FPGA of \lr{the} SCIP are insufficient \lr{for storing} the LUT.
Thus, the bit-compression function is computed pixel-wise using the FPGA resources directly. 
We employ a simple and efficient algorithm to calculate the square root function in Eq.(2) using a \lr{coordinate rotation digital computer} (CORDIC) approach.
The following calculation is implemented on the FPGA for an input value greater than $N_c$:
\begin{eqnarray}
X' &=& \frac{a' + cordic\_sqrt(b'N - c')}{2^5}, 
\end{eqnarray}
where the new constants (\lr{$a'$, $b'$, and $c'$}) are in fixed-point representation and $cordic\_sqrt$ is a module that computes the square root in fixed-point precision using a CORDIC library{\footnote{\linkable{https://www.xilinx.com/products/intellectual-property/cordic.html}}}. 
The parameters for the bit-compression hardware implementation are displayed in Table~\ref{tab:bit_compression}. 
The SCIP data processing pipeline is configured in three different modes: no bit compression, bit compression for unsigned images, and \lr{bit compression} for signed images.
Figure~\ref{fig:LUT} shows the bit-compression functions computed on the FPGA. 

\begin{table}[ht]
\caption{Bit-compression parameters implemented on an FPGA for SCIP} 
\label{tab:bit_compression}
\begin{center}       
\begin{tabular}{|l|l|c|c|c|c|l|} %% this creates two columns
%% |l|l| to left justify each column entry
%% |c|c| to center each column entry
%% use of \rule[]{}{} below opens up each row
\hline
\rule[-1ex]{0pt}{3.5ex}  \# & & a$'$ & b$'$ & c$'$ & $N_c$ & Note  \\
\rule[-1ex]{0pt}{3.5ex}  & & (15 bits)& (20 bits)& (30 bits)& &  \\
\hline\hline
\rule[-1ex]{0pt}{3.5ex}  0 & No compression & - & - & - & - & for rapid mode   \\
\hline
\rule[-1ex]{0pt}{3.5ex}  1 & 22U to 16U & 24976 & 1023984 & 1054707814 & 1280 & for Stokes I   \\
\hline
\rule[-1ex]{0pt}{3.5ex}  2 & 21S to 16U & 25359 & 999427 & 1035405312 & 1280 & for Stokes QUV and R   \\
\hline

\end{tabular}
\end{center}
\end{table}

\subsection{Bit-Compression Test Results}
\label{sec:bit_compression_result}

\begin{figure}
\begin{center}
\begin{tabular}{c}
\includegraphics[bb=0 0 510 340, height=10cm,clip=]{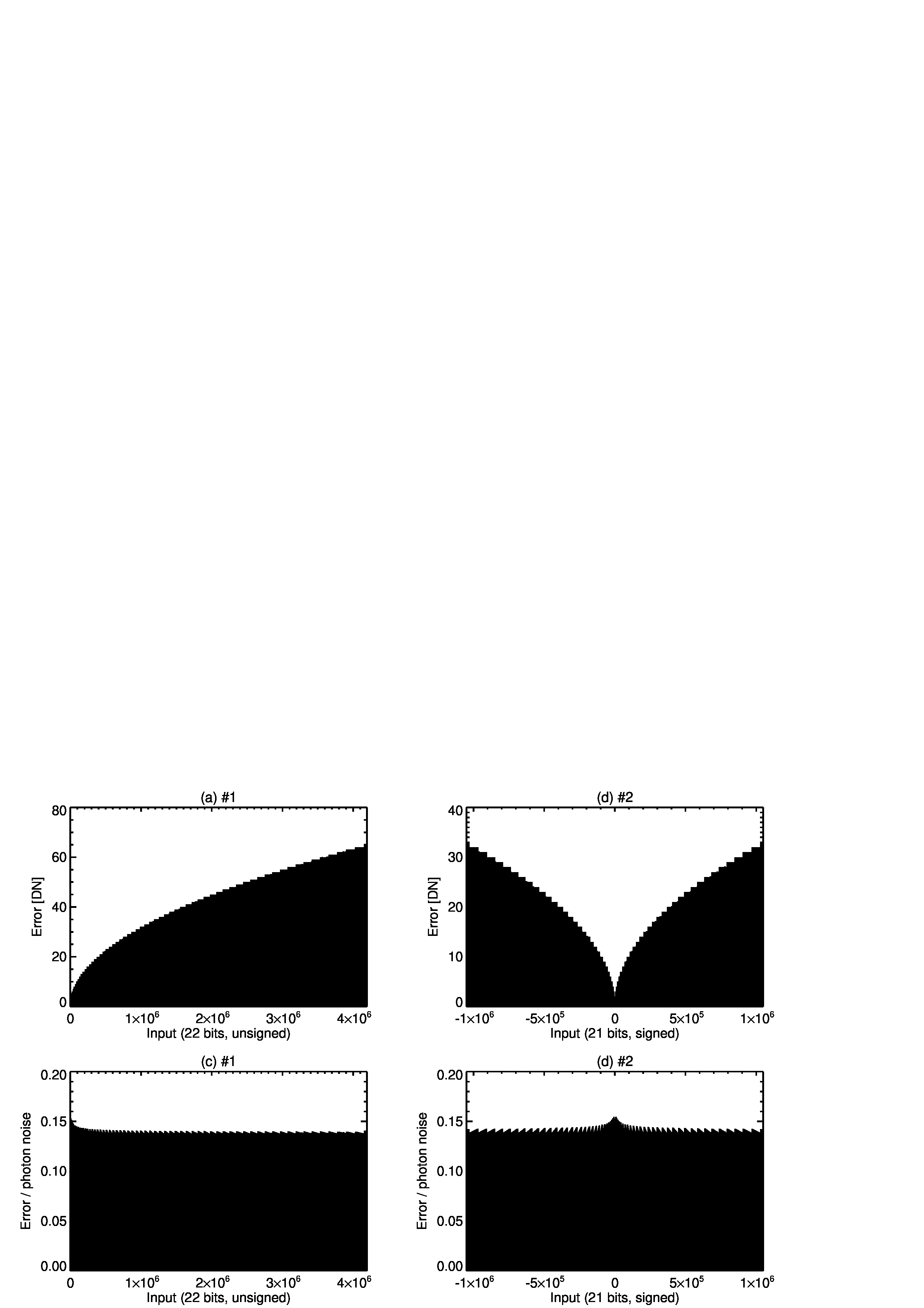}
\end{tabular}
\end{center}
\caption 
{ \label{fig:bitcomp_error}
Bit-compression errors  for (a, c)  Stokes I and (b, d) for Stokes QUV and \lr{the} R-parameter.
The errors are shown in units of DN in the upper panels and relative to photon noise in the lower panels.
}
\end{figure} 

\begin{figure}
\begin{center}
\begin{tabular}{c}
\includegraphics[bb=0 0 453 453, height=16cm,clip=]{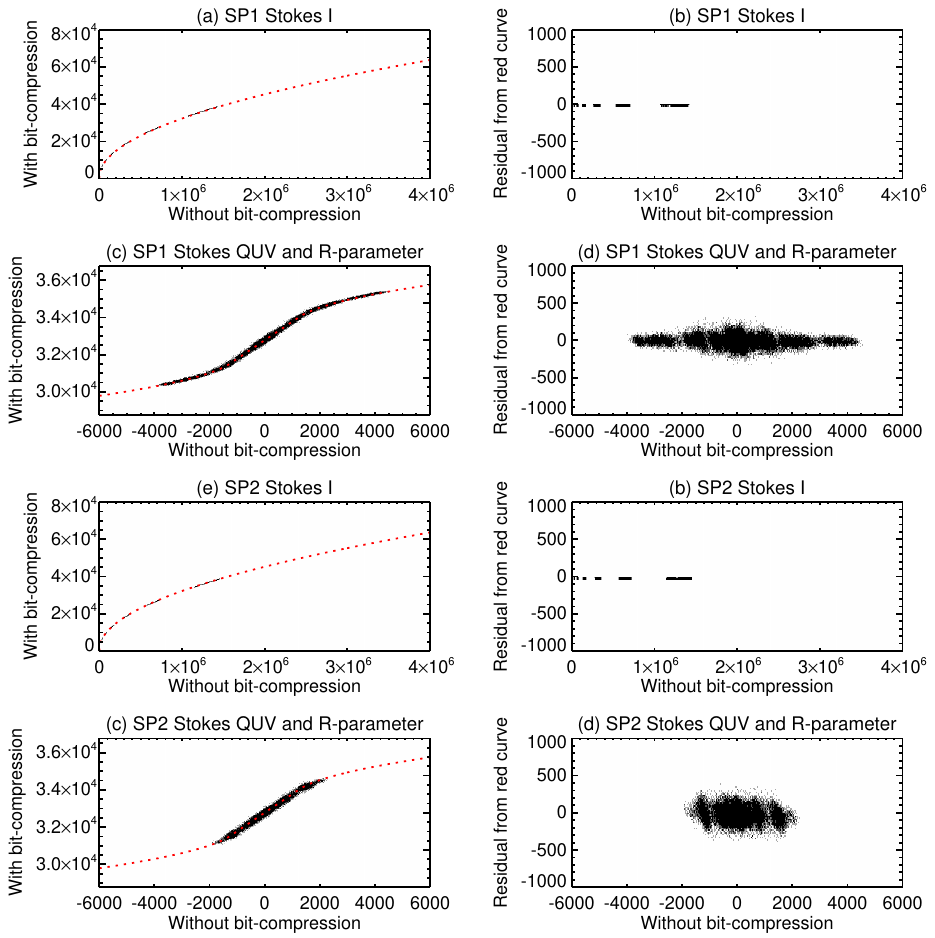}
\end{tabular}
\end{center}
\caption 
{ \label{fig:bitcomp}
(a) Comparison of the observed Stokes I intensity with and without the bit-compression process for SP1 (850 nm). 
The red dashed line is the bit-compression function.
(b) Residual of the Stokes I intensity observed with bit compression from the ideal bit-compression function implemented on an FPGA. 
(c) and (d) are same as (a) and (b), respectively, but for Stokes QUV and \lr{the} R-parameter.
Panels (e) - (h) are the same as the top four panels, but are for SP2 (770 nm).
}
\end{figure} 

The bit-compression process is a lossy and irreversible method, and it introduces errors.
\lr{The} SCIP has a function to generate a dummy image containing the bit compression function.
Using this function, we apply the bit-compression process to the reference sets of the entire input ranges for both unsigned (\#1) and signed (\#2) cases.
The errors are calculated as the difference between the reference set and the results after bit decompression, as shown in Fig.~\ref{fig:bitcomp_error}. 
These errors are compared with photon noise, which is calculated as the square root of the reference sets with the camera gain of 0.052 DN/$e^{-}$.
The errors produced by the bit-compression process are five times less than the photon noise (1$\sigma$) for both unsigned (\#1) and signed (\#2) cases.
This implies that the inherent features of the data are intact even with irreversible compression.

For the verification of bit compression with the flight hardware, we sequentially took the images in the \lr{normal} mode with and without bit compression at different integration times.
In this test, the slit was uniformly illuminated by a white light LED.
Because the intensity of the LED was lower than that of natural sunlight, the maximum integration time of 163.84 s was greater than the nominal range.
The maximum intensity reached slightly higher than 20 bits.
Figure~\ref{fig:bitcomp} shows a comparison of the data taken with and without the bit-compression process.
The bit-compressed data follow the ideal bit-compression functions for both SP channels.
The deviation from the bit-compression functions is mainly caused by the temporal variation between the data with and without bit compression.

\section{Image Compression}
\label{sec:image_compression}

\subsection{Image Compression Algorithm}
\label{sec:image_compression_algorithm}

The image compression algorithm is based on the Lossless Data Compression 121.0-B-2 standard proposed by the \lr{consultative committee for space data systems} (CCSDS)\cite{CCSDS2012}. This algorithm is a low-complexity solution for lossless compression of any type of digital data, including 2D images, \lr{that} require a moderate data rate reduction.
It reaches typically compression factors of 1.5 \lr{to} 2, without any loss during compression/decompression. 

The compressor consists of two functional parts, a preprocessor and an adaptive entropy coder, \lr{that} process the input data partitioned in blocks of $J$ samples with a dynamic range of $n$ bits per sample. An uncompressed sample, called the reference sample, is included in intervals of $r$ blocks to enable the decompression and initialize the process. 
For \lr{the} SCIP, the preprocessor is implemented with a unit-delay predictor and the entropy encoder processes the image in blocks of size $J=16$, $n=16$. The reference sample interval is configured to $r=512$\cite{9970132}.

\subsection{Image Compression Results}
\label{sec:image_compression_result}

The compression efficiency of our compression algorithm is first verified with synthetic SCIP data.
Then, the performance of the onboard image compression process is confirmed on ground with the flight hardware.
An advantage of using the synthetic data is to reproduce the fine scale structures such as granulation and magnetic elements in the solar atmospheres.
These fine scale structures can be resolved by \lr{the} SCIP during the balloon flight, but it is difficult to detect them in our laboratories during the  {\sunriseIII} testing on ground because of the poor atmospheric seeing conditions.

\subsubsection{Synthetic data}

\begin{figure}
\begin{center}
\begin{tabular}{c}
\includegraphics[bb=0 0 425 481, height=17cm,clip=]{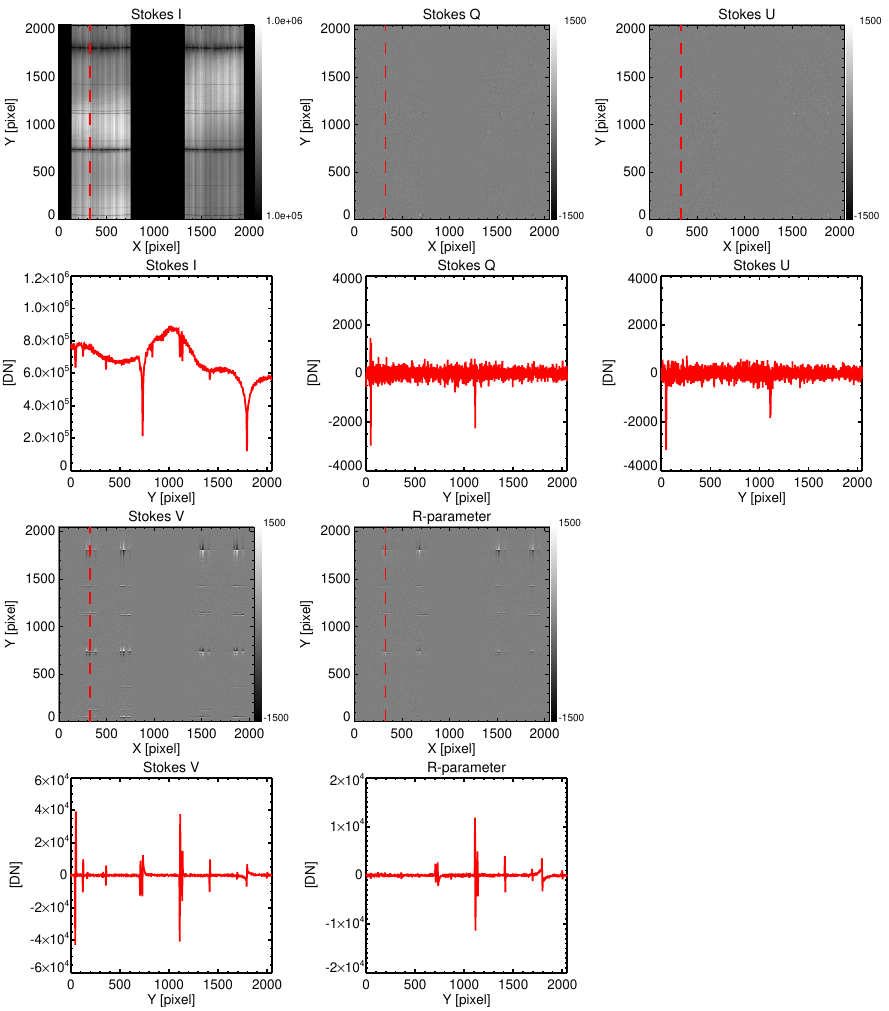}
\end{tabular}
\end{center}
\caption 
{ \label{fig:SP1_sample}
\lr{Synthetic} dataset of SP1 (850 nm) in the \lr{normal} mode. 
The horizontal and vertical axes of the panels in the first and third rows represent the spatial and wavelength directions, respectively.
The panels in the second and fourth rows show the Stokes I, Q, U, V, and R profiles along the vertical dashed line in their upper panels.
The vertical dashed line is located at a small magnetic element with strong field strength.} 
\end{figure} 

\begin{figure}
\begin{center}
\begin{tabular}{c}
\includegraphics[bb=0 0 425 481, height=17cm,clip=]{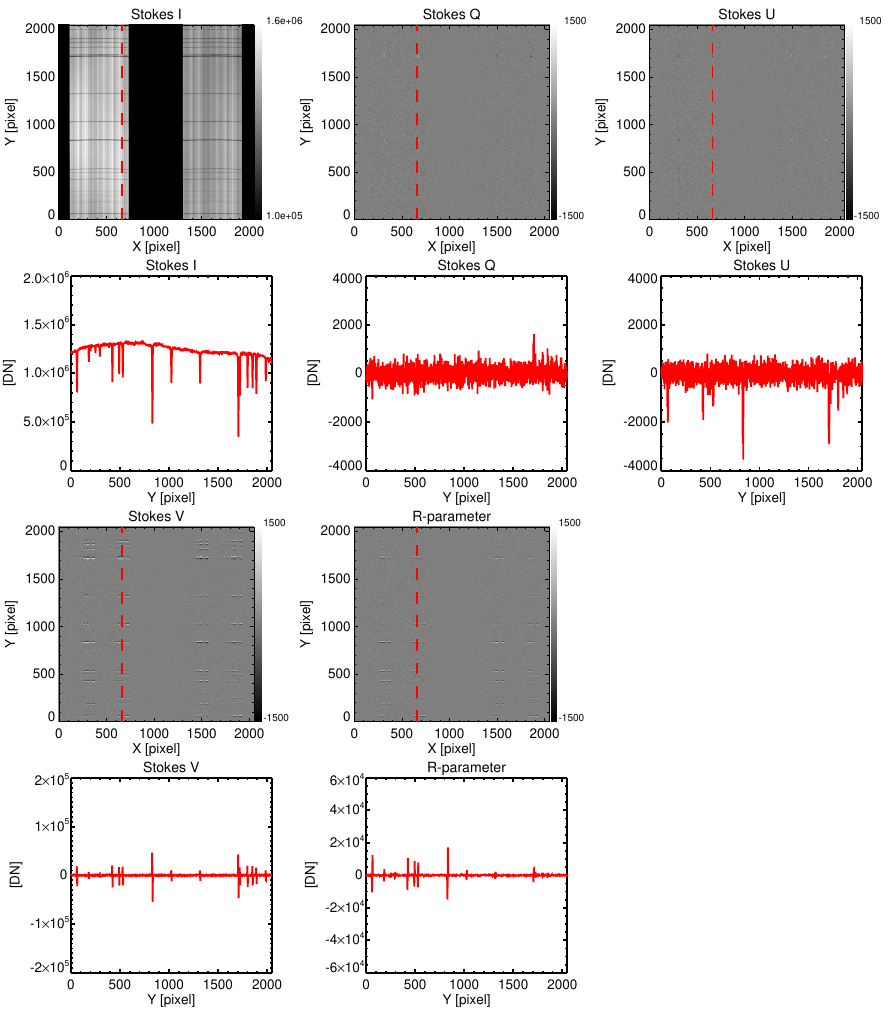}
\end{tabular}
\end{center}
\caption 
{ \label{fig:SP2_sample}
\lr{Synthetic} dataset of SP2 (770 nm) in the \lr{normal} mode. 
The horizontal and vertical axes of the panels in the first and third rows represent the spatial and wavelength directions, respectively.
The panels in the second and fourth rows show the Stokes I, Q, U, V, and R profiles along the vertical dashed line in their upper panels.
The vertical dashed line is located at a small magnetic element with strong field strength}.
\end{figure} 

\begin{figure}
\begin{center}
\begin{tabular}{c}
\includegraphics[bb=0 0 396 113, height=5cm]{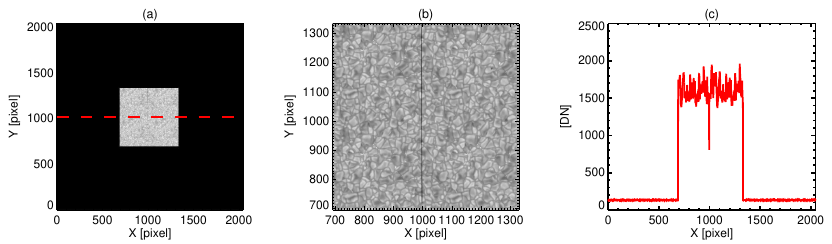}
\end{tabular}
\end{center}
\caption 
{ \label{fig:SJ_sample}
(a) A synthetic SJ image.
(b) Elongated image of the illuminated area in \lr{panel} (a). 
The size of the illuminated area is 640$\times$640 pixels. The central vertical dark line represents the slit.
(c) Intensity profile along the horizontal dashed line in \lr{panel} (a). 
} 
\end{figure}

We use the Rybicki-Hummer (RH) statistical equilibrium and radiative transfer code\cite{2001ApJ...557..389U, 2003ApJ...592.1225U} to synthesize the Stokes profiles.
The geometry package used in RH is the 1-D plane-parallel geometry.
We assume a non-local thermodynamic equilibrium and complete redistribution for computing the atomic populations of K I\cite{1992A&A...265..268U} and the Ca II lines observed with \lr{the} SCIP. 
The atomic information for the different spectral lines can be found in previous studies\cite{2017MNRAS.464.4534Q, 2017MNRAS.470.1453Q}.
We employ an enhanced network simulation\cite{2016A&A...585A...4C} computed with Bifrost code\cite{2011A&A...531A.154G} as solar atmospheres.
The size of the Bifrost simulation is 24 Mm $\times$ 24 Mm on the Sun with a pixel size of 48 km, which corresponds to 0.07 arcsec.
The RH calculations are done with the original resolution of the Bifrost simulation.
The instrumental effects of SCIP are included on the synthetic data as the following steps.
The synthetic SCIP data results are shown in Figs.~\ref{fig:SP1_sample}-\ref{fig:SJ_sample}.

\begin{enumerate}
\item Spatial and spectral degradation

The synthetic SP and SJ data are spatially degraded using a Point Spread Function (PSF).
We assume an ideal PSF corresponding to the \lr{airy} pattern produced by a 1-m telescope.
Spatial binning is done to match the pixel size of \lr{the} SCIP (0.09 arcsec).
\lr{Because} the box size of the Bifrost simulation is smaller than the slit length of 58 arcsec,
periodic boundary conditions are assumed for replicating the field-of-view of the simulation.
The illuminated areas for the orthogonal polarization are embedded in 2k$\times$2k images to keep their positions identical to the ones observed with each SP camera.
The spectral sampling of \lr{the} SCIP is 2$\times$10$^5$, which is 42.5 and 38.4 m{\AA} at the central wavelengths for SP1 and SP2, respectively.
We convolve the full spectrum with the spectral PSF assuming a Gaussian shape to represent instrument degradation along the spectral direction.
The synthetic data contains all of the important absorption lines, although their positions are not perfectly identical to the observed ones because of small errors in the spectral line database.

\item Polarization modulation

To demonstrate the onboard demodulation, the modulated intensity ($I'(\lambda)$) is calculated from the synthetic Stokes \lr{I, Q, U, and V} maps.
We assume the ideal case of PMU: the retardation of the waveplate is 127 \lr{deg} without internal reflection and the PMU phase angle takes discrete values with a step difference of 22.5 \lr{deg}.
\lr{The} phase shift due to the rolling shutter is also considered for the calculation of the modulated intensity.

\item Intensity scaling

The intensity of the SP channels in DN units is calculated as
\begin{equation}
I'_{DN}(\lambda) = I'(\lambda)\,T_h\,G\,t_{exp}.
\end{equation}
The intensity of the original synthetic Stokes profiles is normalized by the local continuum value of each spectral window. 
The expected throughput ($T_h$) for disk center observations is 1.15$\times$10$^6$ $e^{-}$ pixel$^{-1}$ s$^{-1}$ and 2.10$\times$10$^6$ $e^{-}$ pixel$^{-1}$ s$^{-1}$ for the SP1 (850 nm) and SP2 (770 nm) channels, respectively\cite{2020SPIE11447E..0YK}.
The exposure time per frame ($t_{exp}$) is 32 and 10 ms in the \lr{normal and rapid} modes, respectively.
We use a default gain (G) of 0.052 DN/$e^{-}$.
%In the case of SJ,  the exposure time is always 10 ms without the integration. 
The average intensity of the synthetic SJ image is scaled to 1500 DN.

\item Photon noise

We compute the photon noise as the square root of the total number of electrons for the modulated intensity ($I'(\lambda)\,T_h$) at each pixel.
The calculated photon noise is added to Stokes IQUV and R-parameter.

\item Demodulation

We apply a demodulation process similar to that onboard (\lr{Sec.}~\ref{sec:demodulation}).
Datasets with integration of 32 images (1.024 s) and 320 images (10.24 s) were prepared for testing the image compression.

\item Flat field

The synthetic demodulated data were multiplied by a flat field to add the observed patterns into the synthetic images.
The flat field was created from the dataset obtained with the real hardware in the \lr{normal} mode with an integration time of 10.24 s when the SJ field-of-view including the slit was almost uniformly illuminated by the white light LED\cite{2022SPIE12184E..27K}.
After the dark subtraction and gain correction, the Stokes I map was smoothed with a width of 3$\times$3 pixels in the illuminated areas to reduce noise.
The smoothed image \lr{was} normalized by the intensity averaged over the illuminated areas.
The non-illuminated area \lr{was} set to one in the flat-field image.
The SP and SJ synthetic images were multiplied by corresponding flat-field images.

\item Dark 

The measured dark images were added into Stokes \lr{parameter maps} and R maps.
The integration time of the dark images was the same as that for the maps of Stokes parameters.
As a result, the final products of the synthetic SP data contained readout noise, dark current, and bias noise.
The synthetic SJ images also sum with the measured dark images in the same way.

\end{enumerate}

The synthetic Stokes I map for SP1 (Fig.~\ref{fig:SP1_sample}) looks similar to the observed one (Fig.~\ref{fig:SP1_obs}).
However, the bias signals of the synthetic Stokes Q and U data are much smaller than the observed ones, and they fluctuate around zero.
The large signals in the observed maps are polarization induced by the instrument, which is not considered in the synthetic data.
On the other hand, large spiky signals can be seen in the synthetic Stokes V and R-parameter data.
These signals are related to the fine-scale magnetic elements, which cannot be observed with \lr{the} SCIP on the ground because of the poor seeing conditions.

The bit compression and image compression, similar to the onboard ones, are applied to the final products of the synthetic SP data  in the \lr{normal} mode.
For the synthetic SJ data, only image compression is applied.
The results of the compression efficiency in units of bits/pixel are summarized in Table~\ref{tab:img_compression}.
The compression efficiencies match the values used for the prediction of the data rate (``\lr{presumed} values'' in Table~\ref{tab:img_compression}).
The results with \lr{a} larger integration have slightly larger values of bits/pixel but are still smaller than the expected values.
The size of the compressed images for SJ is smaller than those for SP because its bit depth (12 bits) before the image compression is smaller than those of SP (16 bits).
We can assume that the size of the compressed images for the SP channels in the \lr{rapid} mode is similar to the SJ result in Table~\ref{tab:img_compression} because the exposure time and bit depth are the same.

\begin{table}[ht]
\caption{Summary of image compression results in \lr{norma}l mode} 
\label{tab:img_compression}
\begin{center}       
\begin{tabular}{|l|l|l|l|l|l|} %% this creates two columns
%% |l|l| to left justify each column entry
%% |c|c| to center each column entry
%% use of \rule[]{}{} below opens up each row
\hline
\rule[-1ex]{0pt}{3.5ex}  Cameras &Integration& & Presumed values & Synthetic data & Data with flight hardware \\
\rule[-1ex]{0pt}{3.5ex}  & & & [unit: bits/pixel] & [unit: bits/pixel] & [unit: bits/pixel] \\
\hline\hline
\rule[-1ex]{0pt}{3.5ex}  SP1 &1 s& Stokes I & 10 & 7.2 & 7.5     \\
\rule[-1ex]{0pt}{3.5ex}  & & Stokes QUVR & 10 & 7.4 & 7.8    \\
\cline{2-6}
\rule[-1ex]{0pt}{3.5ex}  &10 s& Stokes I & 10 & 8.7 & 8.7     \\
\rule[-1ex]{0pt}{3.5ex}  & & Stokes QUVR & 10 & 9.1 & 9.1    \\
\hline
\rule[-1ex]{0pt}{3.5ex}  SP2 &1 s& Stokes I & 10 & 7.6 & 7.4   \\
\rule[-1ex]{0pt}{3.5ex}  & &Stokes QUVR & 10 & 7.8 & 8.3    \\
\cline{2-6}
\rule[-1ex]{0pt}{3.5ex}  &10 s& Stokes I & 10 & 9.0 & 8.8   \\
\rule[-1ex]{0pt}{3.5ex}  & &Stokes QUVR & 10 & 9.2 & 9.4   \\
\hline
\rule[-1ex]{0pt}{3.5ex}  SJ &10 ms&  & 8 & 5.8 & 7.4   \\
\hline
\end{tabular}
\end{center}
\end{table}

\subsubsection{Data with flight hardware}

\begin{figure}
\begin{center}
\begin{tabular}{c}
\includegraphics[bb=0 0 510 283, height=10cm]{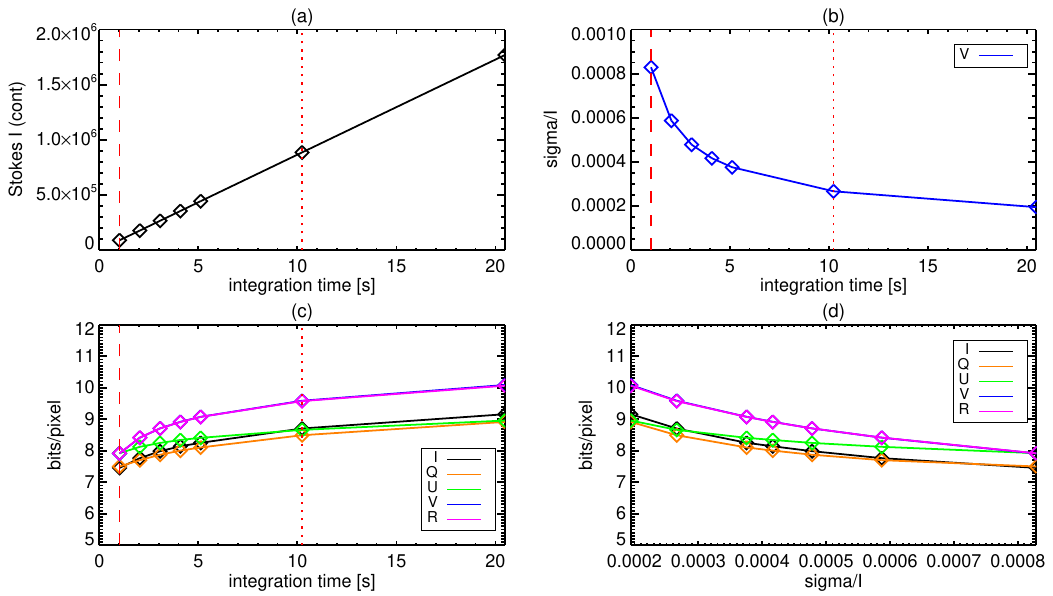}
\end{tabular}
\end{center}
\caption 
{ \label{fig:imgcomp_sunlight}
Results of the natural sunlight test for SP1 (850 nm) in \lr{normal} mode. 
(a) Continuum intensity of Stokes I as a function of integration time.
The continuum intensity in units of DN is averaged over the area without the absorption lines. 
(b) Standard deviation of Stokes V normalized by the continuum intensity as a function of the integration time.
(c) Size of the compressed images as a function of the integration time and (d) the standard deviation of Stokes V normalized by the continuum intensity.
} 
\end{figure} 

We took the datasets for SP1 in the \lr{normal} mode with bit compression and image compression enabled during the sunlight test.
The set of the Stokes maps was obtained at six different integration times from 1.024 to 10.24 s.
The result at the 10.24 s integration is shown in Fig.~\ref{fig:SP1_obs}.
It is confirmed that the onboard processing will allow \lr{the} SCIP to achieve 0.1\% and 0.03\% sensitivities with \lr{1 s} and 10 s integration times, respectively (Fig.~\ref{fig:imgcomp_sunlight}(b)).
The compression efficiency for 0.03\% sensitivity is \lr{$\sim$}9 bits/pixel for Stokes IQU and 10 bits/pixel for Stokes V and the R-parameter (Figs.~\ref{fig:imgcomp_sunlight}(c) and (d)).
These values are slightly larger than those obtained using the synthetic data but are consistent with the presumed ones.
The datasets that were useful for the verification of the image compression were not obtained for SP2 and SJ during the sunlight test because of technical problems.
\lr{Because} the opportunity of the sunlight test is limited, we verified the compression efficiency using the images illuminated by the white light LED.
The integration time for SP2 and exposure time for SJ are adjusted to obtain the number of photons corresponding to those with natural sunlight.
As summarized in Table~\ref{tab:img_compression}, their compression efficiencies are similar to the presumed values as well as SP1.

\section{Conclusion}
\label{sec:summary}
We have verified with the SCIP flight hardware that the performance of the onboard data processing satisfies the required data reduction.
The maximum data rate in the \lr{normal} mode is estimated to be 512 Mbits/s in total.
In this estimation, the integration time of \lr{normal} mode is 1.024 s, and we assumed 10 and 8 bits/pixel for outputs from the SP and SJ cameras, respectively.
The data rate in \lr{rapid} mode was 601 Mbits/s in total with the assumption of 8 bits/pixel for all three cameras.
The data rate in both observing modes has \lr{a sufficient} margin for the limitation of 1 Gbits/s owing to the \lr{gigabit ethernet} connection with the data storage.
Onboard demodulation reduces the number of post-processing methods, \lr{so} the polarization induced by the instruments must be more precisely controlled to achieve precise polarization measurements.
Nevertheless, the onboard demodulation with bit compression and image compression can significantly reduce the telemetry rate.
For example, for 10.24 s integration time, the data rate is reduced by a factor of 77 in the SCIP case: 320 images with 12 bits/pixel \lr{versus} 5 images with 10 bits/pixel.
This reduction is essential to \lr{obtaining} 2048 pixels in the wavelength direction.
Note that the wide wavelength coverage is one of the key advantages in the SCIP observations \lr{for obtaining} three-dimensional magnetic field structures from the solar photosphere to chromosphere.
Another approach to reduce the data rate is to simply sum the equivalent \lr{eight} states from the half rotations of the PMU.
The advantage of this approach is that the full polarimetric correction can be done on the ground.
However, the data rate becomes higher than that of the demodulation into the \lr{five} states.
The number of the output images is increased from 5 to 8.
The intensity of the polarization-modulated images does not change much because of the weak polarization signal.
A larger bit depth per pixel is required for all 8 images when the integration time becomes long.
On the other hand, the bit depth in the Stokes QUV and R-parameter images can be reduced by calculating the difference of the modulated images.

The speed of the onboard processing is also important.
As mentioned in \lr{Sec.}~\ref{sec:intro}, fast modulation and demodulation are necessary for polarization measurements to investigate the rapidly changing dynamics in the solar chromosphere.
The SCIP electronics unit can process 2k$\times$2k images at 31.25 Hz from two cameras. 
Generally, the data rate is considered more seriously in satellite missions than in balloon experiments. 
The high-speed onboard processing of SCIP will be important for future spectropolarimetric observations in satellite missions.

\subsection* {Code, Data, and Materials Availability} 
The data that support the findings of this article are not publicly available due to the data policy of the {\sunriseIII} team. The data can be requested from the authors.

\subsection* {Acknowledgments}
The authors are grateful to two anonymous reviewers for their comments \lr{that improved} the manuscript.
The balloon-borne solar observatory {\sunriseIII} is a mission of the Max Planck Institute for Solar System Research (MPS, Germany), and the Johns Hopkins Applied Physics Laboratory (APL, \lr{United States}). 
{\sunriseIII} looks at the Sun from the stratosphere using a 1-meter telescope, three scientific instruments, and an image stabilization system. Significant contributors to the mission are a Spanish consortium, the National Astronomical Observatory of Japan (NAOJ, Japan), and the Leibniz Institute for Solar Physics (KIS, Germany). The Spanish consortium is led by the Instituto de Astrof\'{i}sica de Andaluc\'{i}a (IAA, Spain) and includes the Instituto Nacional de T\'{e}cnica Aeroespacial (INTA), Universitat de Val\`{e}ncia (UV), Universidad Polit\'{e}cnica de Madrid (UPM) and the Instituto de Astrof\'{i}sica de Canarias (IAC). 
Other partners include NASA's Wallops Flight Facility Balloon Program Office (WFF-BPO) and the Swedish Space Corporation (SSC). {\sunriseIII} is supported by funding from the Max Planck Foundation, NASA \lr{(Grant No. 80NSSC18K0934)}, Spanish FEDER/AEI/MCIU (\lr{Grant No.} RTI2018-096886-C5) and a ``Center of Excellence Severo Ochoa'' award to IAA-CSIC (\lr{Grant No.} SEV-2017-0709), and the ISAS/JAXA Small Mission-of-Opportunity program and JSPS KAKENHI \lr{(Grant No. JP18H05234)}, and NAOJ Research Coordination Committee, NINS. We would also like to acknowledge the technical support from the Advanced Technology Center (ATC), NAOJ. 
We would like to thank Editage (www.editage.com) for English language editing.
%%%%% References %%%%%

\bibliography{report}   % bibliography data in report.bib
\bibliographystyle{spiejour}   % makes bibtex use spiejour.bst

%%%%% Biographies of authors %%%%%

%\vspace{2ex}\noindent\textbf{First Author} is an assistant professor at the University of Optical Engineering. He received his BS and MS degrees in physics from the University of Optics in 1985 and 1987, respectively, and his PhD degree in optics from the Institute of Technology in 1991.  He is the author of more than 50 journal papers and has written three book chapters. His current research interests include optical interconnects, holography, and optoelectronic systems. He is a member of SPIE.
%

\vspace{1ex}
%\noindent Biographies and photographs of the other authors are not available.
\noindent Biographies of the authors are not available.

%\listoffigures
%\listoftables

\end{spacing}
\end{document}